\documentclass{tlpcorr} \usepackage[latin1]{inputenc}
\usepackage{graphicx}

\usepackage{aopmath}

\usepackage{url}

\newcommand{\setlog}{$\{log\}$}

\newcommand{\fourdots}{\mbox{~::~}}

\newcommand{\schit}{\ensuremath{{\mathcal I} \mbox{-Set}}}

\newcommand{\sch}{\iset}

\newcommand{\schs}{\sch s}

\newcommand {\fd}{{\it FD}}

\newcommand{\iset}{\mbox{I-Set}}

\newcommand{\isets}{\iset s}

\newtheorem{definition}{Definition}[section]

 \submitted{31 August 2002}

 \revised{13 November 2003, 30 June 2004}

 \accepted{9 August 2004}

\begin{document}

\title[A CHR-based Implementation of Known Arc-Consistency] {A
  CHR-based Implementation\\ of Known Arc-Consistency}
\author[M.Alberti et al.]  {
  MARCO ALBERTI, MARCO GAVANELLI, EVELINA LAMMA\\
  Dipartimento di Ingegneria, Università degli Studi di Ferrara \and
  PAOLA MELLO, MICHELA MILANO\\
  Dipartimento di Elettronica, Informatica e Sistemistica, Università
  degli Studi di Bologna }

\maketitle
\begin{abstract}
  
  In classical CLP(\fd) systems, domains of variables are completely
  known at the beginning of the constraint propagation process.
  However, in systems interacting with an external environment,
  acquiring the whole domains of variables before the beginning of
  constraint propagation may cause waste of computation time, or even
  obsolescence of the acquired data at the time of use.
  
  For such cases, the Interactive Constraint Satisfaction Problem
  (ICSP) model has been proposed \cite{IJCAI99} as an extension of the
  CSP model, to make it possible to start constraint propagation even
  when domains are not fully known, performing acquisition of domain
  elements only when necessary, and without the need for restarting
  the propagation after every acquisition.
  
  In this paper, we show how a solver for the two sorted CLP language,
  defined in previous work \cite{TOPLAS} to express ICSPs, has been
  implemented in the Constraint Handling Rules (CHR) language, a
  declarative language particularly suitable for high level
  implementation of constraint solvers.

\end{abstract}

\section{Introduction}
Constraint Logic Programming on Finite Domains (CLP(\fd)) represents
one of the most successful implementations of declarative languages.
By means of constraints, the user can give the specifications of a
combinatorial problem and possibly solve it, exploiting efficient
propagation algorithms. CLP(\fd) languages have been successfully used
for solving a variety of industrial and academic problems.  However,
in some constraint problems, where domain elements need to be
acquired, it may not be wise to perform the acquisition of the whole
domains of variables before the beginning of the constraint
propagation process.  For instance, in configuration problems
\cite{DomWildcard,ILOG-Configurator} domain elements represent
components, which have to be synthesized before being used. The set of
components is not known beforehand, and sometimes even the size of the
set cannot be estimated. Often, a minimization of the set of components
is required, thus the constraint solver produces a new component only
when it is strictly necessary.

In systems that need to interact with an external environment, domain
elements can be produced by an acquisition system that retrieves
information about the outer world. An example is given by Faltings and
Macho-Gonzalez \shortcite{Faltings2003} where Internet applications
are faced and obviously not all the information can be computed before
starting the constraint satisfaction process. As another example,
consider a visual search system \cite{PACLP99} where domain elements
are basic visual features (like segments, points, or surface patches)
extracted from the image.  In a classical CLP(\fd) computation, all
domain values must be known when defining the variables, so all the
possible visual features would have to be extracted before starting
the visual search process, even if only a small subset of them will be
actually used. The synthesis of visual features is usually very time
consuming, because the information encoded with signals must be
converted into symbolic form.  Thus, the extraction of domain elements
that will not be used can result in a significant waste of computation
time.  Also, in systems that interact with an evolving environment,
full acquisition of all the domain elements is not wise \cite{ecp99}.
In fact, if all the possible information is acquired beforehand, some
of the information might be obsolete at the end of the acquisition.

For all these reasons, a new model called Interactive Constraint
Satisfaction Problem (ICSP) has been proposed \cite{IJCAI99} as an
extension of the widely used Constraint Satisfaction Problem (CSP)
model. In an ICSP, domains consist of a known part, containing the
available elements, plus a variable that semantically represents a set
of values that could be added to the domain in the future. In a sense,
in an ICSP, domains can be considered as streams of information from
one system to the constraint solver. Constraint propagation can be
performed even when the domains are not completely known, and domain
values can be requested from an acquisition system during constraint
propagation; in other words, constraint propagation and value
acquisition \emph{interact} (thus \textit{Interactive} in the name of
the framework) and are interleaved, whereas in classical CSP
frameworks domain elements are completely known before the beginning
of the propagation.  In this way, the acquisition system can possibly
extract only elements consistent with the imposed constraints, thus
focusing the attention only on significant data.  Various propagation
algorithms have been proposed \cite{NGC2001} for exploiting the
available information and acquiring new domain values only when
strictly necessary.  Reducing the number of extracted elements can
provide a notable speedup \cite{IJCAI99}.

In \cite{TOPLAS} we describe a corresponding CLP language.  Our
language is two sorted. The first sort is the classical sort on Finite
Domains (\fd).  The second sort, called \schit, is based on a
structure similar to {\em streams}, and represents domains of the \fd
\ variables. From the constraints on the \fd \ sort, the system can
start propagation before having full knowledge of domain elements.
Each element will be inserted {\em on demand} in the domain, without
having to restart constraint propagation from scratch.  Moreover,
constraints can be imposed on domains, thus helping the user defining
\isets\footnote{In this article, the \schit\ notation (with
  calligraphic $\mathcal{I}$) will denote the sort, while the notation
  \iset\ (with non-calligraphic I) will denote a particular \iset.}
declaratively.  In this paper, we present an implementation of the
two-sorted language in Constraint Handling Rules (CHR).  CHR
\cite{CHR} is a declarative language for defining new constraint
solvers at a very high level.  CHR can be used for rapid prototyping,
and has proven effective in various real life applications.  The
purpose of this paper is to show how Constraint Handling Rules can be
effectively used to implement the solver for our two sorted ICSP-based
language.  In previous work \cite{IJCAI99}, the propagation algorithms
have been proposed and separately implemented, but we did not describe
the full implementation of the solver for ICSP problems.  The
algorithms were implemented using non fully declarative constructs
(e.g., metaterms with destructive assignment). The high level,
declarative encoding in CHR consists of a solver for the \schit \ 
sort, a solver for the \fd \ sort, and an interface between them
designed to exploit the advantages of the ICSP model in systems
interacting with external acquisition modules.

Of course, other aspects in the model of a problem, besides domain
elements, could be unknown: in a CSP there could be unknown variables
or unknown constraints.  Typically, in all Constraint Programming
systems new constraints can be easily added, while removal of
constraints is more complex \cite{DCSP-Dechter}. The addition of
variables has been taken into account by Dynamic CSP models
\cite{DCSP}. Our work is focussed on unknown domain elements, and
proposes an interaction based on the acquisition, from an external
system, of domain elements.

The rest of the paper is organized as follows.  The declarative and
operational semantics of the language defined by Gavanelli \emph{et
  al.} \shortcite{TOPLAS} are briefly recalled in Section
\ref{syntax}. In Section \ref{sec:impl_concepts} we describe the
architecture of the language from an implementation viewpoint, and in
Section \ref{chr_impl} we show how it is implemented in the CHR
language.  Discussion of related work (including a detailed comparison
with \cite{DomWildcard}, which is very related to our work from the
operational viewpoint) and conclusions follow.

\section{Syntax and  Semantics}
\label{syntax}

The language defined in \cite{TOPLAS} is based on a two sorted CLP,
where the first sort is the classical sort on Finite Domains (\fd) and
the second is the sort on \schit. \isets\ are used both as domains for
\fd \ variables and as communication channels with an external source
providing elements.
                                
In this section, we briefly recall the syntax and semantics of the
language.

In the following, we comply to the conventions in
\cite{semantics-CLP}. In particular, every constraint domain ${\cal
  C}$ (where ${\cal C}$ can be \fd, \schit, \ or \fd+\schit) contains:
the constraint domain signature $\Sigma_{\cal C}$, the class of
constraints ${\mathcal L}_{\cal C}$ (a set of first-order
$\Sigma$-formulas), the domain of computation ${\mathcal D}_{\cal C}$
(a $\Sigma$-structure that is the intended interpretation of
constraints), the constraint theory ${\mathcal T}_{\cal C}$ (a
$\Sigma$-theory that describes the logical semantics of the
constraints), and the solver $solv_{\cal C}$.

\subsection{The \schit \ sort}
\label{syntax-streams}
\label{sec:sch_op_sem}

The \schit\ sort is meant to provide domains for variables in the \fd\ 
sort.  Domains are thus considered as first-class objects, and they
can be declaratively defined by means of constraints. In a sense, they
can be considered as {\em streams}, but they are intrinsically
non-ordered, and do not contain repeated elements.  Declaratively,
\isets\ are sets; thus unification and constraints should consider
\schit \ terms modulo the set theory \cite{LOG}: $\{ A | \{ A | B \}
\} = \{ A | B \}$, $\{ A | \{ B | C \} \} = \{ B | \{ A | C \} \}$,
which states that sets do not contain repeated elements and order is
not important. Non-ground elements are forbidden in an \iset; this
restriction can be exploited for more efficient propagation
algorithms. In fact, we represent an \sch\ as the union of a set of
ground elements (which we name the \emph{known part} of the \sch) and
a variable representing its \emph{unknown part}.

In CLP(\schit), the constraint domain signature, $\Sigma_{\schit}$,
contains the following constraints:
\begin{itemize}
\item {\em s}-{\em member}$(E,S) \Leftrightarrow E \in S$, 
\item {\em union}$(A,B,C) \Leftrightarrow A \cup B = C$,
\item {\em intersection}$(A,B,C) \Leftrightarrow A \cap B = C$,
\item {\em difference}$(A,B,C) \Leftrightarrow A \setminus B = C$,
\item {\em inclusion}$(A,B) \Leftrightarrow A \subseteq B$,
\end{itemize}
where $E$ represents a ground term, and $A$, $B$, and $C$ represent
\isets.

The operational semantics of the \schit\ sort is defined in terms of
{\it state} of \isets, {\it primitives} used to check and modify the
state, and {\it events} over \isets.

The state of an \iset\ is defined by its {\it known part}, i.e. the
{\it set} of the elements which are known to belong to the \iset\ (as
opposed to its {\it unknown part}, representing the elements which
have not yet been acquired for the \iset), and its {\it open-closed}
condition, i.e., an \sch \ is \emph{open} if new elements can be
inserted into it, \emph{closed} otherwise.

A convenient notation to express the state of an \sch \ is one based
on Prolog-like lists. An \sch \ is represented by a structure $S$
defined by $S::=\{\}$, or $S::=\{T|S\}$, or $S::=V$, where $T$ is a
ground term and $V$ is a variable.

The known part of an \sch \ is the set of all the ground elements in
the list representing it; an \sch \ is closed if its continuation
(tail) is ground, open otherwise.  For example, \sch\ $\{1,2,3,4|T\}$
is open, and its known part is the set $\{1,2,3,4\}$; \sch\ 
$\{1,2,3,4\}$ has the same known part, but it is closed.

We have primitives to modify and check the state of an \sch. Two
primitives are used to modify the state, namely:
\begin{itemize}
\item {\it ensure\_member(Element,Iset)}: enforces {\it Element} to be
  a member of the known part of {\it Iset}, possibly adding it if it
  is not already member, or failing if it is not already member and
  the \sch\ is closed;
\item {\it close(Iset)}: closes {\it Iset}; after execution of this
  primitive, no new elements can be added to the \sch.
\end{itemize}
The following primitives are used to check the state of an \sch:
\begin{itemize}
\item {\it known(Iset,KnownPart)}: {\it KnownPart} is the list of
  elements in the known part of {\it Iset};
\item {\it is\_closed(Iset)}: checks if the {\it Iset} is closed.
\end{itemize}
An {\it event} is a notification of how the status of the computation
has been modified, which may be relevant for the rest of the
computation and is to be processed properly. The semantics of an event
is defined by rules specifying its interactions with the constraints
in the store. It is quite apparent that the concept of event finds a
natural representation as a CHR constraint; nevertheless, we prefer,
in defining the operational semantics of the language, to keep events
and proper \schit\ constraints distinct.  The concept of event makes
it possible to express the semantics of \schit \ constraints with
simple (CHR-like) rules. For example, this rule is all that is needed
to define the fact that, in the {\it inclusion}/2 constraint, all
elements in the first \sch\ also appear in the second:
$$\begin{array}{l} inserted(Element,Iset1),inclusion(Iset1,Iset2)
  \Longrightarrow \\
  ensure\_member(Element,Iset2)
\end{array}$$
This rule simply states that, if the \textit{inserted(Element,Iset1)}
event is raised (i.e., if {\it Element} has been inserted into {\it
  Iset1}) and \textit{Iset1} is known to be included in {\it Iset2},
the propagation process must make sure that the element is also
present in {\it Iset2}. This may imply the insertion of {\it Element}
into {\it Iset2}, with a subsequent {\it inserted(Element,Iset2)}
event, if {\it Iset2} is open and {\it Element} is not already a
member, or a failure, if {\it Iset2} is closed and {\it Element} is
not already a member.

Propagation of closure can also be managed with ease. For instance,
the rule
\begin{equation}
\begin{array}{l}
 closed(Iset2), inclusion(Iset1,Iset2),  \\
 known(Iset1,K1), known(Iset2),K2 \Longrightarrow \\
  permutation(K1,K2) | \\
 close(Iset1)
\end{array}
\label{eq:inclusion_closure}
\end{equation}
states that if $Iset1 \subseteq Iset2$, $Iset2$ is closed (as
indicated by the \textit{closed(Iset2)} event), and the
known elements in $Iset1$ are all the known elements in $Iset2$, then
also $Iset1$ is closed (by the \emph{close(Iset1)} primitive).
\subsection{The \fd \  sort}
\label{oper-fd}
The \fd\ sort shares the same declarative semantics of the classical
\fd \ sort. Thus, the usual constraints in CLP(\fd) are considered
(arithmetic, relational constraints plus user-defined constraints). We
suppose that the symbols $<,\leq,+,-,\times,\dots$ belong to
$\Sigma_{\fd}$ and are interpreted as usual.

Since we want cope with incompletely specified variable domains
avoiding useless value acquisition, we use a constraint propagation
based on the available knowledge, when domains are still partially
specified. For this reason, we proposed \cite{TOPLAS} an extension,
for the partially known case, of the concept of consistency, called
{\em known consistency}. In this paper, we provide only the definition
of node and arc-consistency; the extension to higher degrees of
consistency is straightforward.

\begin{definition}
  A unary constraint $c(X_i)$ is {\em known node-consistent} iff
  $$\forall v_i \in K^c_i, \ v_i \in c(X_i),$$
  where $K^c_i$ is the
  known part of the domain of $X_i$.  A binary constraint $c(X_i,X_j)$
  is {\em known arc-consistent} iff
  $$\forall v_i \in K^c_i , \ \exists v_j \in K^c_j \ \emph{s.t.} \ 
  (v_i,v_j) \in c(X_i,X_j),$$
  where $K^c_i$ and $K^c_j$ are the known
  parts of the domains of $X_i$ and $X_j$, respectively.
  \label{defKAC} A constraint network is {\em known arc-consistent
    (KAC)} iff all unary constraints are known node-consistent and all
  binary constraints are known arc-consistent.
\end{definition}

The following proposition shows the link between \textit{known
  arc-consistency} and \textit{arc-consistency} \cite{Mac77}.  The
proof can be found in \cite{TOPLAS}.
\begin{proposition}
  \label{prop} Every algorithm achieving KAC (i.e. any algorithm
  that computes an equivalent problem that is KAC) and that ensures at
  least a known element in each variable domain is able to detect
  inconsistency in the same instances as an algorithm achieving AC.
\end{proposition}
In other words, if there exists an arc-consistent sub-domain, then
there exists a maximal arc-consistent sub-domain; so if KAC does not
detect inconsistency, AC will not detect inconsistency either.

KAC is equivalent to AC when domains are completely known. The
advantage in using KAC is that the check for known arc-consistency can
be performed lazily, without full knowledge of all the elements in
every domain.

\subsection{\label{sec:linking-two-sorts}Linking the two sorts}
\label{sec:link-sorts}
Intuitively, we want to bind CLP(\fd) with CLP(\schit) with the
intended semantics that \sch s provide domains for \fd\ variables.

Given the two CLP languages ${\cal L}_{\fd}$ and ${\cal L}_{\schit}$,
we define the CLP language ${\cal L}$ as the union of the two
languages, with a further constraint,~$\mbox{~::~}$, defined as
follows:
 \begin{itemize}
 \item the signature $\Sigma = \Sigma_{\fd} \cup \Sigma_{\sch} \cup
   \{\mbox{~::~}\}$;
 \item the intended interpretation ${\cal D}$ keeps the original
   mappings in the \fd \ and \schit\ sorts; i.e., ${\cal
     D}|_{\Sigma_{\fd}}={\cal D}_{\fd}$ and ${\cal
     D}|_{\Sigma_{\sch}}={\cal D}_{\sch}$.
\end{itemize}
The declarative semantics of the constraint is
$$X \mbox{~::~} S \leftrightarrow X \in S$$
where $X$ is a \fd\ 
variable. The \fourdots/2 constraint links a \fd\ variable to its
domain as in most $CLP(FD)$ frameworks, with the difference that $S$,
being and \iset, may be non-completely specified. The \fourdots/2
constraint should not be confused with the {\em s}-{\em member}/2
constraint of Section \ref{syntax-streams}, which represents the
constraint of set membership between a ground element and an \iset.

In CLP(\fd) systems, domains provide ancillary information about
variables. Domains contain the possible values that a variable can
take; if a value is not consistent with the imposed constraints, it is
operationally removed from the domain.  This helps many systems
\cite{CHIP,ILOG,eclipse5.2,SICStus} to obtain higher performance; in
fact, domain wipe-outs are detected early and many alternatives are
efficiently pruned.  On the other hand, in the \schit \ sort, domains
must be manipulated as logical entities: if an element declaratively
belongs to a domain, it cannot be removed.

Suppose that we have constraint $\{1,2,3\} \subseteq D$ stating that
the elements 1, 2 and 3 should belong to the set $D$, the constraint
$X \fourdots D$ that links variable X to the domain D, and $X \neq 1$
that states that X should be different from 1.  Usual constraint
propagation of the constraint $X \neq 1$ would remove element 1 from
the domain $D$, but this is inconsistent with the constraint
$\{1,2,3\} \subseteq D$ so the computation would fail. This behavior
is not correct, because the set of constraints $\{1,2,3\} \subseteq
D$, $X \in D$, and $X \neq 1$ is satisfiable.

For this reason, in our framework, the domain of a variable $X$ is
represented by two streams: a {\em Definition Domain}, $D^d_X$, that
contains all the values synthesized for the variable, and a stream of
{\em Removed Values}, $D^r_X$, that is the set of elements proven to
be inconsistent with the imposed constraints.  The set of available
items for $X$, also called {\em Current Domain}, $D^c_X$, is given by
the relation:
                                \begin{equation}
  D^c_X = D^d_X \setminus D^r_X.\label{vincolo_dom}$$
\end{equation}
It should be noticed that $D^c_X$ remains open until $D^d_X$ is
closed. $D^r_X$, instead, is always open, so to make it possible to
move elements into it from the current domain even if the definition
domain has been closed.
                                
\begin{figure}
  \includegraphics[width=6cm]{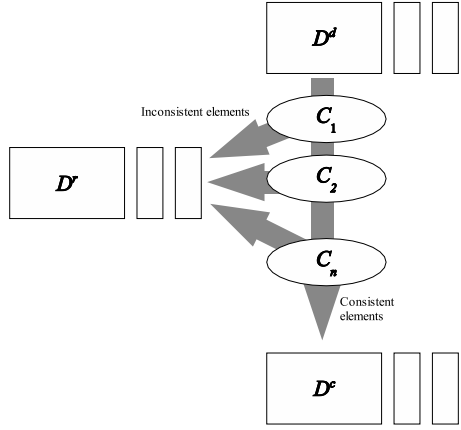}\\
  \caption{FD Constraints as filters on variable domains}
  \label{filters}
\end{figure}

With the imposed constraints, it is possible to declaratively add a
newly synthesized value to the definition domain (imposing the
constraint $v \in D^d_X$) or remove an inconsistent element from the
current domain ($w \notin D^c_X$ or, equivalently, $w \in D^r_X$; see
Fig.  \ref{filters}). In this way, during search, an inconsistent
element will not be tried if it belongs to $D^r_X$ and a possible
domain wipe-out will be detected when $D^c_X$ is empty.

Note that the relation in Eq. (\ref{vincolo_dom}) automatically
propagates two types of information from the definition domain to the
current domain of a variable. First, whenever a domain element is
synthesized, it is considered in the current domain and can be
exploited for propagation. Second, if no more values are available to
the definition domain, also the current domain becomes closed.

Notice that more than one FD variable can range on the same definition
domain or, equivalently, their definition domains can be linked by an
equality constraint; however, each of them will have its own current
domain and set of removed values.  By definition, the user cannot
close the current domain of a variable: the user should only access
the definition domain directly. This is not a restriction, because if
one wants to close independently the current domain of different
variables (as in the previous example), he can define two different
definition domains (and, possibly, impose some constraints among them,
e.g., $D^d_X \subseteq D^d_Y$).
  
In order to achieve KAC, it is necessary to remove elements and to
{\em promote} elements, i.e., to move ideally some elements from the
unknown part to the known part. Elements can then be removed (i.e.,
prevented from entering the current domain) if they are shown to be
inconsistent. An algorithm for achieving KAC is shown in
\cite{TOPLAS}.

The addition, besides the deletion, of elements to the domain might
seem non monotonic. However, the current domain is kept open as long
as new acquisitions are possible (i.e., until the definition domain is
closed), the unknown part representing the set of future acquisitions.
Thus, declaratively, when a new element enters the known part of the
current domain, it is not properly added: it is just made explicit. A
similar behavior is also achieved by Mailharro \shortcite{DomWildcard}
with the use of a \emph{wildcard} representing values {\em entering}
the domain.
  
Clearly, the repeated acquisition of the same element would result in
a loop; this can be avoided by having an acquisition module that does
not provide twice the same element to the same \sch\ (as hypothesized,
for instance, by Mailharro \shortcite{DomWildcard}).

\paragraph{Example: Numeric CSP.}
\label{sec:examples}                    \label{numeric}
    
Various applications could be thought exploiting interactive
constraint propagation, as many systems need to interact with an
external environment during propagation: some real-world examples can
be found in \cite{TOPLAS}. In this section, we give a simple example,
aimed at showing a typical ICSP computation, rather than at showing
the power of the language.

With the given language, we can state in a natural way the following
problem:
$$\mbox{\tt :-}X\mbox{~::~}D_X, Y\mbox{~::~}D_Y, Z\mbox{~::~}D_Z,
intersection(D_X,D_Y,D_Z), Z>X.$$
defining three variables, $X$, $Y$,
and $Z$, with constraints on them and their domains.  KAC propagation
can start even with domains fully unknown, i.e., when $D_X$, $D_Y$ and
$D_Z$ are variables. Let us suppose that the elements are acquired
through interaction with a user and the first element retrieved for
$X$ is $1$. This element is inserted in the definition domain of $X$,
i.e., $D^d_X = \{1 | (D^d_X)'\}$; since it is consistent with FD
constraints, it is not redirected to the $D^r_X$ stream; value 1 is
considered in the current domain, i.e., $D^c_X = \{1 | (D^c_X)'\}$.
Then KAC propagation tries to find a support for this element in each
domain of those variables linked by \fd \ constraints; in our instance
$D_Z$.  A value is requested for $D_Z$ and the user gives a (possibly
consistent) value: $2$. This element is inserted into the definition
domain of $Z$: $D^d_Z=\{2 | (D^d_Z)'\}$. The constraint imposed on
domains can propagate, so element $2$ has to be inserted in the
definition domain of $Y$ and $X$, thus $D^c_Y=D^d_Y=\{2|D'_Y\}$ and
$D^c_X=D^d_X=\{1,2|D''_X\}$.

Since the acquired element is consistent with FD constraints involving
$Z$, it enters the current domain of $Z$.  KAC propagation must now
find a known support for element 2 for $X$, so another request is
performed. If the user replies that there is no other element in the
domain of Z, then $2$ is sent to the stream of removed values for $X$
(thus, it is not considered in the current domain of $X$). It will
remain in the definition domain because the element semantically
belongs to the domain even if no consistent solution can exist
containing it.

When KAC propagation reaches the quiescence, each element of the
current domain of each variable has a support for each FD constraint
involving that variable.

\section{Implementation concepts}
\label{sec:impl_concepts}

Since one of the aims of the ICSP model is to manage efficiently those
problems in which value acquisition is costly, it is useful to exploit
the link between the two sorts to avoid unnecessary acquisition of
domain elements.  For instance, as shown in the example in Section
\ref{numeric}, it is possible to infer the presence of a value in the
known part of an \sch\ from \schit\ constraints, without acquiring it
directly.

Thus, for the sake of efficiency, the implementation of the language
must be aware of the link between the two sorts, given by the
\fourdots/2 constraint (see Section \ref{sec:linking-two-sorts}), and
exploit it when useful.  Nonetheless, the two sorts are clearly
distinct, and need to be handled by different computational
mechanisms.

In more detail, it is possible to devise two constraint solvers: one
for the \fd\ sort, meant to ensure KAC (see Def. \ref{defKAC}) of the
\fd\ constraint network, and one for the \schit\ sort, meant to ensure
satisfaction of \schit\ constraints.

These solvers only need to interact in two cases:
\begin{itemize}
\item when, during \fd\ KAC check, a new element is acquired, \schit\ 
  propagation must be activated in order to satisfy \schit\ 
  constraints;
\item conversely, when a new element enters the known part of an \sch,
  \fd\ KAC check must be activated over all the \fd\ variables which
  have their definition domain in the \sch.
\end{itemize}
The CHR language is a perfectly suitable tool for this purpose,
letting us deal with the two sorts separately, and to manage the link
between them with ease.

\subsection{\fd\  sort concepts}

The aim of the \fd\ solver is to keep the constraint network known
arc-consistent.  This is achieved by preventing elements from entering
current domains if no KAC support is found for the constraints
involving them.  In more detail, each time a new element $E$ is
promoted from the unknown to the known part of an \sch\ $C$, it is
inserted into the definition domain of each variable $V$ such that
$V\fourdots C$ ($E \in D^d_V$ with the notation of Sect.
\ref{sec:link-sorts}). The algorithm then tries to find a KAC support
for $E$: if support is found, then $E$ is also inserted into the {\it
  Current Domain} of $V$ ($E \in D^c_V$); otherwise, although it stays
in the definition domain of $V$, it enters the stream of {\it Removed
  Values} for $V$ ($E \in D^r_V$).

In this way, at any step of the propagation, all of the values in
current domains are certainly supported.
\subsubsection{\label{sec:values-states}Values' states}

\begin{figure}
    \includegraphics[width=6 cm]{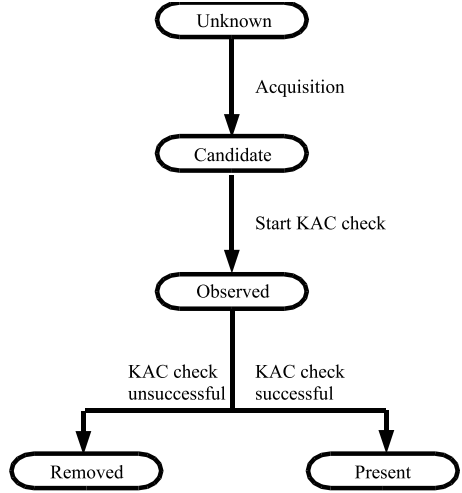}\\
    \caption{State transitions for ({\it Variable},{\it Element})
      pairs}\label{fig:states}
  \end{figure}

The process described above can be clearly formalized as a sequence of
state transitions for $(V,E)$ pairs.  These states are:

\begin{itemize}
  
\item{{\it unknown}}: $E$ has not yet been acquired as a value for
  $V$;
  
\item{{\it candidate}}: $E$ has been inserted into $D^d_V$, but
  $(V,E)$ has not yet been chosen for supporting another value and no
  KAC control is being run on it;
  
\item{{\it observed}}: $(V,E)$ provides support for some other {\it
    observed} pair, but it has not yet been proven to be supported;
  
\item{{\it present}}: $(V,E)$ is supported, and has thus been inserted
  into $D^c_V$;
  
\item{{\it removed}}: $E$ is not supported, and has thus been inserted
  into $D^r_V$.

\end{itemize}
The possible state transitions are shown in Figure \ref{fig:states}.

\subsubsection{Lazy value acquisition}

The ICSP framework is more suitable for those applications in which
the process of acquiring domain values is computationally costly.  In
this perspective, the KAC check procedure has been designed so as to
acquire new values only when it is really necessary, i.e., when no
support for a value can be found among already known values.

\subsection{\label{sec:sch-sort-concepts}\schit\ sort concepts}

The task performed by the \schit\ solver is to ensure the satisfaction
of the \schit\ sort constraints, and to propagate the closure of
definition domains, when possible. As shown in the example below, it
is possible to infer that an element belongs to an \sch\ from \schit\ 
sort constraints, thus reducing the number of necessary acquisitions
and improving efficiency.

In particular, each time an element is added to an \iset, an
\textit{inserted}/2 event (see Section \ref{syntax-streams}) is
raised, which starts a check for the satisfaction of the \schit\ 
constraints involving the \iset\ itself.

\paragraph {Example}

Let us suppose the constraint {\it intersection}($D_X$,$D_Y$,$D_Z$) is
given among the $D_X$, $D_Y$ and $D_Z$ \schs, and that the current
values of the domains are

$D_X=\{2,4|D_X'\}$; $D_Y=\{3,4|D_Y'\}$; $D_Z=\{4|D_Z'\}$.

\noindent If $5$ is added to $D_Z$, the system immediately ensures $5$ to be a
member of both $D_X$ and $D_Y$, without the need for acquiring the
element.  If 3 is added to $D_Z$, the system only needs to add it to
$D_X$. Conversely, if 3 is added to $D_X$, it is also added to $D_Z$;
if 1 is added to $D_X$, instead, the system cannot infer anything, and
thus makes no additions.

\schit\ constraints also have to be taken into account to propagate
closure of \sch s, when possible (as in the example shown in Sect.
\ref{sec:sch_op_sem}). For this purpose, we use a \emph{closed}/1
event, which notifies that an \sch\ has been closed.
\subsection{Interaction between sorts}

The link between the two sorts, represented by the \fourdots /2
constraint (see Section \ref{sec:linking-two-sorts}), is implemented
as an interaction mechanism between the solvers activated when, on the
one hand, a new element is acquired during a KAC check and, on the
other hand, an element is inserted into an \sch.

Semantically, we use the {\it inserted}/2 event for both of these cases
(see Section \ref{sec:sch_op_sem}).

At the implementation level, the event causes the control to pass from
one solver to another.

\subsection{\label{algorithm}Algorithms}

\subsubsection{\label{sec:prior-among-supp}Priority among supporting values}

A {\it (Variable,Element)} pair is supported if, for each constraint
involving {\it Variable}, support is found in the known part of other
variables' domains. But supporting pairs may not be supported in their
turn: this implies that, if a pair is found to be unsupported, new
support needs to be found for all of the pairs that it was supporting.
For efficiency, we keep track of which pairs support which other pairs
for which constraints, so that, in case a pair is found unsupported,
it is possible to revise only the affected constraints.

However, because of states (see Sect. \ref{sec:values-states}), it is
possible to reduce the number of supports to keep track of.  If an
element is supported by an element whose state is \emph{present},
there is no need for keeping track of the support, because a present
element is already known to be supported, as explained in Sect.
\ref{sec:values-states}.  Thus, when seeking support for a constraint,
present elements should be tried first.

Likewise, an element whose state is \emph{observed} is more convenient
than an element whose state is \emph{candidate}, because the algorithm
is already seeking support for it, and can thus avoid seeking support
for a new element (note that all candidates will eventually be
checked; but in this way, when their turn comes, there will be, in
general, more present elements).

So, a priority can be established among eligible supporting elements:
first present, then observed, then candidates. Only when no support is
found among these elements, a new element needs to be acquired: if the
definition domain of the corresponding variable is open, an element
can be requested, otherwise the support seek procedure fails.

\subsubsection{\label{subsubsec:support_graph}The support graph}

Support dependencies can be represented by a (directed) support graph,
in which nodes represent observed {\it (Variable,Element)} pairs, and
arcs represent support dependencies.

When a candidate pair becomes observed, the corresponding node is
added to the graph. This may happen at the beginning of graph
construction, when the graph is empty and the candidate is chosen to
be the first node, or later, during the support seeking procedure, if
the candidate provides support for an observed.

A new node needs support for the \fd\ constraints involving it. For
each \fd\ constraint, the following attempts, in this order (see Sect.
\ref{sec:prior-among-supp}) are made, until one succeeds:
\begin{enumerate}
\item support is found in present elements only: the graph needs no
  modifications;
\item one of the supporting elements is observed: an arc is added from
  the supporting node to the supported, marked with the constraint;
\item one of the supporting elements is candidate: a new node is added
  for the candidate, an arc is added from the supporting node to the
  supported, marked with the constraint.
\end{enumerate}
If none of these attempts succeeds, a variable (other than
\textit{Variable}) is chosen for acquisition among those involved by
the constraint, and the search for support is restarted from the
support-in-candidates attempt (step 3).  The search for support needs
to be restarted because, in order to satisfy \schit\ constraints, new
candidates may have been added, which may provide support for the
considered \fd\ constraint.

A node is deleted from the graph if it is found to have no support; in
this case, new support is sought for all of the nodes that it was
supporting.

Once a graph is fully built, all of its nodes can be inserted into the
current domains (observed-to-present transition), whereas the
unsupported nodes will be inserted into the stream of removed values.

The procedure reaches quiescence only when there are no more
candidates to search support for.

\section{\label{chr_impl}CHR implementation}

In this section, we show how the concepts explained in Section
\ref{sec:impl_concepts} have been implemented in a constraint solver
using the Constraint Handling Rules library of SICStus Prolog.

\subsection{\label{sec:sch--objects}\schit\ variables and constraints}

\schs\ are represented by CHR variables: these variables (also ``link
variables'' hereafter) act as a link for the information stored in
constraints, and are never instantiated.  To keep memory of the state
of \sch s we use CHR constraints, meant to interact in order to ensure
satisfaction of \schit\ constraints.

\sloppy

The state of an \sch\ is represented by means of three CHR
constraints: {\it iset\_known(Iset,Known)} links an \iset\ to its
known part, {\it iset\_open(Iset)} indicates that the \iset\ is open
(i.e., other elements can be inserted), and {\it closed(Iset)}
indicates that the \iset\ is closed.

\fussy

An \sch\ is created by the user either explicitly ({\it
  new\_iset\_object}/3 predicate) or implicitly, imposing constraint
{\it icsp\_def\_domain}/2 (which implements the \fourdots/2 constraint of
Sect. \ref{sec:linking-two-sorts}) between an \fd\ variable and the
link variable of an \sch.  It is possible to define the domain as
empty or as having a starting known part.

\sloppy

Each constraint among \sch s (see Section \ref{syntax-streams} for the
formal definitions) is represented by a corresponding CHR constraint
among the link variables of the \sch s that it involves, namely the
{\it iset\_member}/2 (in this case, the first argument is a ground
\fd\ value), {\it iset\_union}/3, {\it iset\_intersection}/3, {\it
  iset\_inclusion}/2, and {\it iset\_difference}/3 constraints.

\fussy

The satisfaction of \schit\ constraints is ensured incrementally (see
Section \ref{sec:sch-sort-concepts}).  For instance, the following
rule is used for implementing the {\it iset\_intersection}/3
constraint.

\begin{verbatim}
intersection_right_to_left @
        iset_inserted(Iset3,Element),
        iset_intersection(Iset1,Iset2,Iset3)
                        # _iset_intersection
        ==> ensure_membership(Iset1,Element),
        ensure_membership(Iset2,Element)
        pragma passive(_iset_intersection).
\end{verbatim}
\sloppy

The rule is activated when a new element is added to the \sch\ 
represented by the third argument of an {\it iset\_intersection}/3
constraint: this is notified by the \textit{iset\_inserted}/2
constraint, which implements the \textit{inserted} event of Sect.
\ref{sec:sch_op_sem}. The rule imposes constraint {\it
  ensure\_membership}/2 on each of the other two \isets\ and the
element; if necessary, the element will be inserted into the \isets.
Notice that {\it ensure\_membership}/2 is defined so as to fail if the
\sch\ is closed and the element is not already in its known part.
Constraint {\it iset\_intersection}/3 is declared passive for
efficiency, because it is not necessary to generate code for it
(\schit\ constraints as \textit{iset\_intersection}/3 are imposed only
at the beginning of the computation).

\fussy

Two more rules (not shown here for lack of space) manage the case of
an element having been inserted into the first or second argument of
the {\it iset\_intersection}/3 constraint.

The \emph{close}/1 primitive of Sect. \ref{sec:sch_op_sem} is
implemented by the following CHR:

\begin{verbatim}
close_iset @
        close(Iset), iset_open(Iset) # _iset_open
        <=> closed(Iset)
        pragma passive(_iset_open).
\end{verbatim}

CHR constraint \emph{close}/1 is imposed to close the \sch; the effect
of this rule is to remove constraint \emph{iset\_open}/1 for the \sch,
and to impose CHR constraint \emph{closed}/1 for the \sch.  Constraint
\emph{closed}/1 also implements the event notifying that the argument
\sch\ has been closed, which can interact with \schit\ constraints so
as to propagate closure when possible. For instance, the rule of Sect.
\ref{sec:sch_op_sem} is implemented as follows:

\begin{verbatim}
closure_propagation_inclusion @
        closed(Iset2),
        iset_inclusion(Iset1,Iset2) # _iset_inclusion,
        iset_known(Iset1,K1) # _iset_known1,
        iset_known(Iset2,K2) # _iset_known2,
        ==> permutation(K1,K2) | 
        close(Iset1) pragma passive(_iset_inclusion),
        passive(_iset_known1), passive(_iset_known2).
\end{verbatim}

\noindent Predicate \emph{permutation}/2 in the guard
checks that the two \sch s have the same elements.

\subsection{\fd\  variables and constraints}

\subsubsection{\fd\  variables and domains}

\sloppy

\fd\ variables are represented as CHR constrained variables. Their
domain is an \sch; constraint \fourdots/2 (see Section
\ref{sec:linking-two-sorts}) is implemented as the {\it
  icsp\_def\_domain(Variable,Iset)} CHR constraint, whereas the
current domain of the variable and its set of removed values are
represented by the {\it
  icsp\_curr\_domain(Variable,(Present,Removed))} CHR constraint,
where {\it Present} is the list of present elements and {\it Removed}
is the list of removed elements.

\fussy

\subsubsection{\label{sec:fd-constraints}\fd\  constraints}

\fd\ constraints are represented by the {\it
  fd\_constraint(ConstraintName, ListOfArguments)} CHR constraint. For
instance, \fd\ constraint $X<Z$ is represented by the
\texttt{fd\_constraint(lt,[X,Z])} CHR constraint.

\fd\ constraints are defined by the user simply as one or more clauses
for the Prolog predicate {\it fd\_verify}/2 needed to verify whether
the constraint is satisfied by a list of ground arguments.  This way
of defining \fd\ constraints supports non-binary constraints, and easy
extensibility of the system.

For instance, a possible definition of the \fd\ $<$/2 constraint could
be

\begin{verbatim}
fd_verify(lt,[A,B]):- A<B.
\end{verbatim}

\subsubsection{KAC procedure}

Candidate (see Section \ref{sec:values-states}) elements for each
variable are stored in the {\it
  current\_candidates(Variable,ListOfElements)} constraint.

KAC check over candidates is performed by building a support graph
(see Section \ref{subsubsec:support_graph}). Nodes of the graph are
{\it (Variable, Element)} pairs. A pair is a node of the graph if and
only if it is member of the list argument of the {\it
  observed\_candidates}/1 CHR constraint; an arc (indicating support
dependency) is represented by the {\it relies ((Var1,Val1),
  (Var2,Val2), FDConstraint)} CHR constraint, where {\it FDConstraint}
is the \fd\ constraint for which {\it (Var2,Val2)} supports {\it
  (Var1,Val1)}.

Graph construction is achieved by the following CHR rule:

\begin{verbatim}
kac_check_start @
        kac_unlock,
        current_candidates(Var,[Candidate|MoreCandidates])
                # _current_candidates
        <=> current_candidates(Var,MoreCandidates),
        observed_candidate((Var,Candidate)),
        check_candidate((Var,Candidate)),
        !, flush_candidates, end_flush_candidates,
        kac_unlock pragma passive(_current_candidates).
\end{verbatim}

\noindent Graph construction begins when there is at least one non-empty list of
candidates for a variable: an element is chosen (in the current
implementation, it is simply the first of the list), pair {\it
  (Variable, Element)} becomes the first node of the graph (constraint
\textit{observed\_candidate}/1 adds its argument to the list argument
of constraint \textit{observed\_candidates}/1), and KAC check starts
from it (\textit{check\_candidate}/2 constraint); during check, new
nodes may be added to the graph, which will be checked for support in
their turn. When the construction is done, graph nodes are inserted
into the current domains (\textit{flush\_candidates}/0 and
\textit{end\_flush\_candidates}/0 constraints), and the procedure can
start again over the rest of candidates.

{\it kac\_unlock}/0 is a constraint meant simply to start the KAC
procedure. If there are no candidates when it is imposed, then if
there is a variable with empty current domain an element is acquired
for it which will become a candidate; otherwise, the procedure reaches
quiescence.

\subsubsection{Support seek}

\sloppy

Support seek for a \emph{(Variable, Element)} pair is started by
constraint {\it check\_candidate((Variable,Element))}. This constraint
collects from the store all the \fd\ constraints that involve {\it
  Variable}, and for each of them tries to find an assignment of all
the involved variables that satisfies it (i.e., that makes goal {\it
  fd\_verify}/2 (see Section \ref{sec:fd-constraints}) succeed).
Elements for the other variables are chosen following the priority
described in Section \ref{sec:prior-among-supp}.

\fussy

\subsubsection{Support seek through element acquisition}

If no acceptable assignment is found among known, observed, and
candidate elements, then:
\begin{itemize}
\item if the \sch\ domain of at least one of the other variables is
  open, a new element is acquired for one of the other variables;
\item if the \sch\ domains of all the other variables are closed,
  failure is reported, and possibly backtracking is applied.
\end{itemize}
Which variable should be chosen for acquisition among those with open
domain is not obvious for non-binary constraints, and application
specific heuristics could be useful; in this implementation the first
variable is simply chosen.

In the current system, element acquisition is obtained by asking the
user for an element, and the \sch\ is closed in case of a predefined
input from the user. Obviously, in practical applications, elements
would be provided by an acquisition system, as in \cite{DomWildcard},
where the acquisition is done automatically by generation of new
component instances, or in \cite{PACLP99}, where elements are provided
by a low-level segmentation system.

\subsection{Constraints and rules for interaction between sorts}
\label{sec:constr-rules-inter}

Only one CHR constraint needs to be added to the solver to link the
two sorts: {\it iset\_inserted(Iset,Element)}, imposed when {\it
  Element} is inserted into the known part of {\it Iset}. This
constraint is the implementation of the {\it inserted}/2 event described
in Section \ref{sec:sch_op_sem}.

\subsubsection{\fd\  to \schit\ link}

When a new element $\mathit{Element}$ is acquired for a variable $V$
whose definition domain is $\mathit{Iset}$, {\it
  iset\_inserted(Iset,Element)} is imposed. This CHR constraint
implements an event that triggers the \schit\ constraint check. From a
procedural point of view, this makes control pass from the \fd \ 
solver to the \schit\ solver.

\subsubsection{\schit\ to \fd\  link}

When imposed, {\it iset\_inserted(Iset,Element)}:
\begin{itemize}
  
\item first, interacts with the \schit\ constraints involving {\it
    Iset}, as shown in Section~\ref{sec:sch--objects};
\item then, inserts {\it Element} in the list of candidates of all the
  variables having {\it Iset} as definition domain, for later support
  seek.

\end{itemize}
The interaction with the \schit\ constraints may involve new
insertions, which will, in their turn, impose further {\it
  iset\_inserted}/2 constraints.

New candidates are added only when \schit\ propagation has been
completed, for the reasons explained in Section
\ref{sec:sch-sort-concepts}.

\section {Related work}
\label{related}

\schs\ can be considered as streams with a set semantics (i.e., in an
\sch\ there are no repeated elements and elements are not sorted).
Streams are widely used for communication purposes in concurrent logic
programming \cite{Shapiro}. Various communication protocols can be
implemented using this simple yet powerful data structure
\cite{Shapiro-survey}. An \sch\ can only contain ground elements; this
restriction prevents (open) \isets\ from being passed in an \iset, but
lets us achieve higher efficiency.

The first results of our research in the ICSP framework are reported
by Cucchiara et al. \shortcite{IJCAI99}, where the ICSP model is
proposed as an extension of the CSP model. In this model, variables
range on partially known domains which have a known part and an
unknown part represented as a variable. Domain values are provided by
an extraction module and the acquisition process is (possibly) driven
by constraints. The model has been proven effective in a vision system
\cite{PACLP99}, in randomly-generated problems \cite{IJCAI99}, and in
planning \cite{ecp99}. This work can be considered as the language
extension and CHR implementation of the ICSP framework, maintaining it
as the core of the propagation engine on the \fd\ side.
  
Operationally, achieving KAC has some similarities with achieving {\em
  Lazy Arc Consistency} (LAC) \cite{LazyAC}. LAC is an algorithm that
finds an arc-consistent sub-domain (not necessarily a maximal one) and
tries to avoid the check for consistency of all the elements in every
domain. KAC looks for an arc-consistent sub-domain as well, but it is
aimed at avoiding unnecessary information retrieval, rather than
unnecessary constraint checks.

\sloppy

Codognet and Diaz \shortcite{codognet96} describe a method for
compiling constraints in CLP(\fd). There is only one primitive
constraint ($X$~{\tt in}~$R$), used to implement all the other
constraints. $R$ represents a collection of objects and can also be a
user function. Thus, in CLP(\fd) domains are managed as first-class
objects; our framework can be fruitfully implemented in systems
exploiting this idea.

\fussy

Sergot \shortcite{sergot} proposes a framework to deal with
interaction with the user in a logic programming environment. Our work
can be used for interaction in a CLP framework; it lets the user
interactively provide domain values.

Zweben and Eskey \shortcite{ZE89} propose an algorithm that evaluates
domain elements only when necessary; domains are streams and
constraints are filters on domains. Dent and Mercer
\shortcite{MinimalFC} show the effectiveness of such an approach when
constraint checks are expensive operations. In our proposal,
implementation of delayed evaluation is quite natural and simple, even
if our work is aimed at reducing domain values extractions, not
constraint checks.

Dynamic Constraint Satisfaction (DCS) \cite{DCSP-Dechter} is a
promising field of AI taking into account dynamic changes of the
constraint store such as the addition and deletion of values and
constraints.  The difference between DCS and our approach concerns the
way of handling these changes. DCS approaches propagate constraints as
if they worked in a {\em closed world}. Basically, in a DCS one can
add or remove a constraint; thus, one can also add and delete domain
elements provided that they are all known from the beginning. In an
ICSP, instead, domain elements that are unknown can be requested and
inserted in the domain. DCS solvers record dependencies between
constraints and the corresponding propagation in proper data
structures \cite{DCSP-Nogood} so as to tackle modifications of the
constraint store as soon as data change.  In this perspective, we also
cope with changes since the acquisition of new values can be seen as a
modification of the constraint store. However, we work in an {\em open
  world} where domains are left {\em open} thanks to their unknown
part.  Unknown domain parts represent intensionally future
acquisitions, i.e., future changes.

\sloppy

Another DCS framework was given for configuration
problems~\cite{DCSP}. This framework considers dynamicity in the set
of variables; variables are introduced or removed during search by
means of constraints. The aim is to find a solution where only some of
the variables are assigned a value, while the others are {\em
  inactive}. However, differently from our framework, the set of
domain values is given at the specification of the problem.

\fussy

Many systems consider implement sets, because sets have powerful
description capabilities. In particular, some have been described as
instances of the general CLP(${\cal X}$) framework
\cite{semantics-CLP}, like \{log\} \cite{DovRos93,LOG,setlog_ngc},
CLPS \cite{CLPS-plilp}, or Conjunto \cite{conjunto}. Others apply an
object-oriented approach \cite{PECOS,ILOG}.

In \setlog \cite{DovRos93,LOG,setlog_ngc}, a set can be either the
empty set $\emptyset$, or defined by a constructor {\tt with} which,
given a set $S$ and an element $e$, returns the set composed of $S
\cup \{e\}$. This language is very powerful, allowing sets and
variables to belong to sets. However, set unification and the
propagation of other constraints has an exponential time complexity in
the worst case. If we allow non-ground elements in an \iset, we obtain
a more expressive (although less efficient) framework, like \{log\};
we are currently studying this extension.

Set variables \cite{PECOS,ILOG} can range on set domains. Each domain
is represented by its greatest lower bound and its least upper bound.
Each element in a set must be ground, sets are finite, and they cannot
contain sets. These restrictions avoid the non-deterministic
unification algorithm, and give good performance results; they are
implemented in most Constraint Programming
systems~\cite{ILOG,Oz,SICStus,conjunto,eclipse5.2}.  However, these
systems do not deal with problems in which some domain elements are
not known; in fact, the user must provide the least upper bound of
each set, thus giving the universe set at the beginning of the
computation.

ILOG-Solver \cite{ILOG} does not deal directly with those problems,
but has an extension module, called ILOG-Configurator
\cite{DomWildcard,ILOG-Configurator} whose main added value is to
implement open domains in a way similar to our system. This system is
very related with ours, thus we give a more detailed comparison.

\paragraph{Comparison with ILOG-Configurator.}
\sloppy

In Ilog-Configurator, there are variables called ``ports'' whose
domains are defined by the set of instances of a given component type.
The set of instances is not known in advance and instances are
generated on demand during constraint propagation. In this way,
domains of port variables are dynamically extended; a domain that can
be extended contains an element called ``wildcard''.

\fussy

Our system has many points in common with the one described by
Mailharro \shortcite{DomWildcard}.  To compare the two, the reader can
refer to the following table:

\noindent
\begin{tabular}{ ll @{}}
 \hline \\
 ICSP & ILOG-Configurator \\
 \hline
 \hline
 FD Variables & Port Variables \\
 \sch & Component Types \\
 Known Part & Set of generated instances \\
 Unknown Part & Set of not yet generated instances (Wildcard) \\
 ::/2 & Link port-type \\
 Definition Domain $D^d_X$ & Set of instances of the target type \\
 Current Domain $D^c_X$ & Possible set of a port \\
 Removed Elements $D^r_X$ & Instances set of the target type $\setminus$
possible set
                                           \\
 Inserted Event & Extension delta domain of ports \\
 \hline
\end{tabular}

The differences that we envisage between the two systems are the
following. ILOG-Configurator is based on an object-oriented
technology, while ours is based on logic programming; this is
reflected in some specific choices made in the two systems. For
example, both the systems reason about the possible closure of a
domain. This information is carried by a special element, called
``wildcard" by Mailharro \shortcite{DomWildcard}, while in our system
it is the unknown part of the domain. In our system, if the
continuation of the domain is a variable, other elements can be added,
otherwise, if it is the empty set, the insertion of other elements
will be forbidden by unification. This has a declarative meaning from
a set viewpoint; while a set containing a wildcard must be updated
through some destructive assignment, our formalization lets the user
specify the value of a logical variable: the continuation of the
domain will be extended, as in a stream. In other words, in the ICSP
formulation we keep set membership and set inclusion distinct. In
fact, in \cite{DomWildcard}, the wildcard element represents a set of
elements, but is also an element of the domain. Thanks to our
formalization, some future extensions are possible: for example, it is
possible to extend the type of sets in a domain, and to have domains
that can contain sets themselves, like in \setlog\ \cite{LOG}.  The
propagation of FD constraints is operationally performed in a similar
way (Known Arc Consistency propagation); however, the property of
Known Arc Consistency (KAC), which was not defined by Mailharro
\shortcite{DomWildcard}, enabled us to prove properties of the
algorithms achieving KAC \cite{TOPLAS}.

Finally, in our framework it is possible to define \schs\ as the
combination of other \schs\ with set operators.

\section{Conclusions}
\label{conclusion}

In this work, we presented the implementation of a language that
performs constraint propagation on variables with finite domains when
information about domains is not fully known, and its CHR
implementation. Domains are channels of information, and are
considered as first-class objects that can be themselves defined by
means of constraints. The obtained language belongs to the CLP class
and deals with two sorts: the \fd \ sort on finite domains and the
\schit \ sort for domains. We provide a propagation engine for the
\fd\ sort exploiting known arc-consistency, and one for the \schit\ 
sort, as well as a mechanism for their interaction.

The source code of the system is available on request.

\section*{Acknowledgements}
\label{sec:acknowledgements}

This work is partially funded by the Information Society Technologies
programme of the European Commission under the IST-2001-32530 project
(SOCS) within the Global Computing initiative.

We wish to thank the anonymous reviewers for their useful comments.

\end{document}